\documentclass[11pt,fleqn]{article}
\usepackage{epsfig}
\def\greaterthansquiggle{\raise.3ex\hbox{$>$\kern-.75em\lower1ex\hbox{$\sim$}}}
\def\lessthansquiggle{\raise.3ex\hbox{$<$\kern-.75em\lower1ex\hbox{$\sim$}}}

\newcommand{\la}{\label}
\newcommand{\re}{\ref}
\newcommand{\ci}{\cite}
\newcommand{\beqn}{\begin{eqnarray}}
\newcommand{\eeqn}{\end{eqnarray}}
\newcommand{\bequ}{\begin{equation}}
\newcommand{\eequ}{\end{equation}}
\newcommand{\bsl}{\begin{sloppypar}}
\newcommand{\esl}{\end{sloppypar}}

\begin{document}
\setlength{\unitlength}{1cm}
\null
\hfill DESY 01-117\\
\null
\hfill WUE-ITP-2001-032\\
\vskip .4cm
\begin{center}
{\Large \bf Influence of CP and CPT on production\\[.4em] 
 and decay of Dirac and Majorana fermions}\\[.4em]
\vskip 1.5em

{\large
{\sc 
G.~Moortgat--Pick$^{a}$\footnote{e-mail:
    gudrid@mail.desy.de},
H. Fraas$^{b}$\footnote{e-mail:
    fraas@physik.uni-wuerzburg.de}
}}\\[3ex]                     
\end{center}
{\footnotesize \it
$^{a}$ DESY, Deutsches Elektronen--Synchrotron, D--22603 Hamburg, Germany\\
$^{b}$ Institut f\"ur Theoretische Physik und Astrophysik, Universit\"at
W\"urzburg, D--97074 W\"urzburg,\\ \phantom{$^{b}$ }Germany}\\
\vskip .5em
\par
\vskip .4cm

\begin{abstract}
The consequences of CP and CPT invariance for production and
subsequent decay of Dirac and Majorana fermions in polarized
fermion--antifermion annihilation are analytically studied.  We derive
general symmetry relations for the production spin density matrix and
for the three--particle decay matrices and obtain constraints for the
polarization and the spin--spin correlations of Dirac and Majorana
fermions. We prove that only for Majorana fermions the energy and
opening angle distribution factorizes exactly into contributions from
production and decay if CP is conserved.
\end{abstract}
\vspace{1em}
\hfill

\newpage
\section{Introduction}
\label{sec:1}

It is generally accepted that supersymmetry (SUSY) is one of 
the most promising concepts for physics beyond the Standard Model (SM).
A special feature of SUSY models is the Majorana character of
the neutralinos, the fermionic superpartners 
of the neutral gauge and Higgs bosons. 
After the observation of candidates for neutralinos 
at a future $e^+e^-$ collider \ci{TDR}, 
their identification as Majorana particles is indispensable.
Therefore an extensive study of the general characteristics of Majorana
fermions produced in $e^+ e^-$ annihilation is of particular interest.

In \ci{Petkov,Christova}, 
the authors analyzed the consequences of CP and CPT invariance 
for the symmetry properties of  
the cross section for Majorana fermions produced in 
$e^+ e^-$ annihilation with polarized beams. 
They proposed useful methods to distinguish between Majorana
and neutral Dirac particles via the energy distributions of the leptons from 
their leptonic decay.

In the present paper we extend the investigation of CP and CPT
symmetry properties to the complete spin production density matrix for
Dirac and Majorana fermions and to their decay matrices for a
three--body decay.  We consider the most general dependence on beam
polarization so that our analysis is also applicable to $\mu^+ \mu^-$
annihilation with the exchange of Higgs bosons.

CP and CPT symmetry relations lead to
important consequences for the facto\-rization of
differential cross sections for production and subsequent decay into a
production and a decay piece. In \ci{Tata} it has been proven that
the differential cross section facto\-rizes only if the kinematic
variables are properly chosen. However, factorization is 
ruled out in particular for the energy
and angular distributions of the decay products in the lab frame
due to interference between the helicity amplitudes.

In numerical analyses \ci{Gudi_for} it was demonstrated that for
production and three--particle decay of neutralinos in $e^+e^-$
annihilation the energy distribution of the decay products in the lab
frame as well as the distribution of the opening angle between two of
them indeed factorize. Contrary, for production and three--particle
decay of charginos \ci{Gudi_talk} the spin correlations between
production and decay 
considerably contribute to the energy
distribution of the decay products and therefore
prevent factorization.  
Here we prove that the factorization
of these observables
in the case of Majorana fermions is a consequence of their CP/CPT
symmetry properties. We show that the analysis of both the
energy spectrum of the decay fermions and the opening angle
distribution is useful for the identification of the Majorana character
of neutral particles produced in fermion--antifermion annihilation with 
polarized beams.

This paper is organized as follows:
In section~2.1 we give details of the spin density matrix 
formalism which is applied to the analysis of the consequences of
CPT and CP invariance for the 
production density matrix of two different Dirac and Majorana 
fermions in section~2.2 and 
for three--body decay matrices in section~2.3. 
In section~3 we study analytically the 
consequences for angular distributions and the energy spectra of the 
decay fermions. 
\section{ Constraints on the production of Dirac\\ \hspace*{.1cm} 
and Majorana fermions from
CPT and CP}\la{sec:3}
\subsection{Spin--density matrix and cross section}\la{sec:2}
We consider pair production of Dirac or Majorana fermions in 
fermion--antifermion annihilation and their subsequent three--particle decay.
The helicity amplitudes for the production processes 
\bequ
f(\vec{p}_1,\lambda_1) \bar{f}(\vec{p}_2, \lambda_2)\to f_i(\vec{p}_i, 
\lambda_i) \bar{f}_j(\vec{p}_j, \lambda_j)
\la{eq2_1}
\eequ
are denoted by $T_{P, \lambda_1 \lambda_2}^{\lambda_i\lambda_j}$. 
In the case of Majorana particles it is $\bar{f}_j\equiv f_j$.
For the decay processes
\beqn
f_{i}(\vec{p}_i, \lambda_i) & \to & f_{i1}(\vec{p}_{i1})f_{i2}(\vec{p}_{i2})
f_{i3}(\vec{p}_{i3}),\la{eq2_2a}\\
f_{j}(\vec{p}_j, \lambda_j) & \to &  f_{j1}(\vec{p}_{j1})f_{j2}(\vec{p}_{j2})
f_{j3}(\vec{p}_{i3})\la{eq2_2b}
\eeqn
the helicity amplitudes are given 
by $T_{D,\lambda_i}$ and $T_{D,\lambda_j}$.
Here
$f_{i1}$ ($f_{j1}$) is a charged or neutral 
Dirac or a Majorana fermion and $f_{i2}$ ($f_{j2}$), $f_{i3}$ ($f_{j3}$)
are  Dirac fermions.
The notation of the decays (\ref{eq2_2a}), (\ref{eq2_2b}) 
does not distinguish
between fermions and antifermions
for the initial particles and their decay products. 
The helicities of the decay products are also suppressed.

The spin density matrices of the polarized beams can be written as 
\beqn
\rho(f) & = & \frac{1}{2}(1+P^i_f\sigma^i) \\
\rho(\bar{f}) & = & \frac{1}{2}(1+P^i_{\bar{f}}\sigma^i), 
\eeqn
where 
$P^1_{f}$, $P^2_f$, $P^3_f$ 
($P^1_{\bar{f}}$, $P^2_{\bar{f}}$, $P^3_{\bar{f}}$)
is the transverse polarization of $f$ ($\bar{f}$) in the production
plane, the polarization which is  perpendicular 
to the production plane and the longitudinal
polarization. 

The amplitude squared of the combined process of production and decay reads
\mathindent0cm
\bequ
|T|^2=\sum_{\lambda_i\lambda_j\lambda_i^{'}\lambda_j^{'}}
|\Delta(f_i)|^2 |\Delta(f_j)|^2 
\rho_P^{\lambda_i \lambda_j, \lambda^{'}_i \lambda^{'}_j} 
\rho_{D, \lambda^{'}_i\lambda_i}\rho_{D,\lambda_j^{'} \lambda_j}.
\la{eq4_4e}
\eequ
It is composed from the (unnormalized) spin density production matrix
\bequ
\rho_P^{\lambda_i\lambda_j, \lambda_i^{'}\lambda_j^{'}}=
\sum_{\lambda_1 \lambda_2 \lambda_1^{'} \lambda_2^{'}}
\rho(f)_{\lambda'_1\lambda_1}\rho(\bar{f})_{\lambda'_2 \lambda_2}
T_{P, \lambda_1 \lambda_2}^{\lambda_i\lambda_j}
T_{P, \lambda'_1 \lambda'_2}^{\lambda_i^{'}\lambda_j^{'}*}, \la{eq_rev1}
\eequ
the decay matrices
\beqn
\rho_{D, \lambda_i^{'} \lambda_i} & = & T_{D, \lambda_i}T_{D,\lambda'_i}^{*}\\ 
\rho_{D, \lambda_j^{'} \lambda_j} & = & T_{D, \lambda_j}T_{D,\lambda'_j}^{*}
\la{eq_rev2}
\eeqn
and the propagators 
\bequ
\Delta(f_{k})=1/[p^2_{k}-m_{k}^2+i m_{k} \Gamma_{k}], \; k=i,j.
\eequ
Here $p_{k}^2$, $m_{k}$ and $\Gamma_{k}$ denote the 
four--momentum squared, mass and total width of the fermion 
$f_{k}$. For these propagators we use the narrow--width 
approximation. 

We introduce for $f_i$ ($f_j$)
three spacelike polarization vectors
 $s^{a\mu}(f_i)$
($s^{b\mu}(f_j)$) which together with $p_i^{\mu}/m_i$
($p_j^{\mu}/m_j$) form an orthonormal set.
Then the spin density matrix of production and the decay matrices can
be expanded in terms of Pauli matrices $\sigma^a$ with the first (second)
row and column corresponding to the helicity $\lambda=\frac{1}{2}$ 
($-\frac{1}{2}$) \ci{Haber}:
\beqn 
\rho_P^{\lambda_i\lambda_j, \lambda_i^{'}\lambda_j^{'}}&=&
\delta_{\lambda_i\lambda_i^{'}} \delta_{\lambda_j\lambda_j^{'}}
P(f_i f_j)
+\delta_{\lambda_j\lambda_j^{'}}\sum_{a=1}^3
\sigma^a_{\lambda_i\lambda_i^{'}}\Sigma^a_P(f_i)
\nonumber\\ &&
+\delta_{\lambda_i\lambda_i^{'}}\sum_{b=1}^3
\sigma^b_{\lambda_j\lambda_j^{'}}\Sigma^b_P(f_j)
+\sum_{a,b=1}^3\sigma^a_{\lambda_i\lambda_i^{'}}
\sigma^b_{\lambda_j\lambda_j^{'}}
\Sigma^{ab}_P(f_i f_j),\la{eq4_4h}\\
\rho_{D,\lambda_i^{'}\lambda_i}&=&\delta_{\lambda_i^{'}\lambda_i}
D(f_i)+\sum_{a=1}^3
\sigma^a_{\lambda_i^{'}\lambda_i} \Sigma^a_D(f_i),\la{eq4_4i}\\
\rho_{D,\lambda_j^{'}\lambda_j}&=&\delta_{\lambda_j^{'}\lambda_j}
D(f_j)+\sum_{b=1}^3
\sigma^b_{\lambda_j^{'}\lambda_j} \Sigma^b_D(f_j).\la{eq4_4j}
\eeqn
Here $a$ $(b)=1,2,3$ refers to the 
polarization vectors of $f_i$ ($f_j$).
The contributions $\Sigma_P^a(f_i)$ ($\Sigma_P^b(f_j)$),
$\Sigma_D^a(f_i)$ ($\Sigma_D^b(f_j)$) are linear, and 
$\Sigma_P^{ab}(f_if_j)$ is bilinear in the polarization vectors 
$s^{a\mu}(f_i)$ ($s^{b\nu}(f_j)$). 
In (\re{eq4_4h}) the dependence of $\rho_P$
on beam polarization has been suppressed.

The polarization vectors 
$\vec{s^1}$, $\vec{s^2}$, $\vec{s^3}$ form an orthogonal right--handed 
system in the lab system:
\begin{itemize}
\item
$\vec{s^{3}}(f_i)$ ($\vec{s^{3}}(f_j)$) is in the direction of 
momentum $\vec{p_i}$ ($\vec{p_j}$), 
\item
$\vec{s^2}(f_i)=\frac{\vec{p}_1\times \vec{p}_i}
{|\vec{p}_1\times \vec{p}_i|}=\vec{s^2}(f_j)$ 
is perpendicular to the production plane, 
\item
$\vec{s^1}(f_i)$ ($\vec{s^1}(f_j)$) is in the production plane 
orthogonal to the momentum $\vec{p}_i$ ($\vec{p}_j$).
\end{itemize}

Then 
$\Sigma^3_P(f_{i,j})/P(f_i f_j)$ 
is the longitudinal polarization,
$\Sigma^1_P(f_{i,j})/P(f_i f_j)$ is 
the transverse polarization in the scattering plane and
$\Sigma^2_P(f_{i,j})/P(f_i f_j)$ 
is the polarization perpendicular to the scattering plane.
The terms $\Sigma^{ab}_P(f_i f_j)$ are due to correlations between 
the polarizations of both produced particles. 

The amplitude squared $|T|^2$  
of the combined process of production and decay (\ref{eq4_4e}) can
be written as:
\beqn
|T|^2&=&4|\Delta(f_i)|^2|\Delta(f_j)|^2\Big[P(f_i f_j) D(f_i) D(f_j)
    +\sum^3_{a=1}\Sigma_P^a(f_i) \Sigma_D^a(f_i) D(f_j)
\nonumber\\&&
+\sum^3_{b=1}\Sigma_P^b(f_j) \Sigma_D^b(f_j)
D(f_i)
    +\sum^3_{a,b=1}\Sigma_P^{ab}(f_i f_j)
 \Sigma^a_D(f_i) \Sigma^b_D(f_j)\Big],
\la{eq4_5}
\eeqn
and the differential cross section is 
\begin{equation}
d\sigma=\frac{1}{2 s}|T|^2 (2\pi)^4
\delta^4(p_1+p_2-\sum_{i} p_i) d{\rm lips}\label{eq_13},
\end{equation}
with the Lorentz invariant phase space element $d{\rm lips}$.

For the case of neutralinos and charginos    
the complete analytical expressions for 
the production density matrix and for the decay matrices 
are given in different presentations in \ci{Gudi_for,Choi}.
\subsection{ CPT and CP symmetries of the spin--density matrix}\la{sec:30}
In this section we derive constraints from CPT and
CP invariance for the density matrix (\re{eq4_4h}) for the production of
two (charged or neutral) Dirac fermions and of two different Majorana
fermions, respectively, for polarized beams. 

All contributions from the exchange of particles $\alpha$,
$\beta$ to the coefficients 
\bequ
{\cal E}=\{P, \Sigma^a_P, \Sigma^b_P,
\Sigma^{ab}_P\} 
\eequ
of the spin--density matrix (\ref{eq4_4h}) are
composed of pro\-ducts ${\cal C}$ of couplings, the propagators
$\Delta(\alpha)$, $\Delta(\beta)$ and complex functions ${\cal S}$ of
momenta and polarization vectors: 
\bequ
{\cal E}\sim Re\{{\cal C} \times
\Delta(\alpha) \Delta(\beta)^{*} \times {\cal S} \}.
\eequ
Assuming
complex couplings and taking into account the finite widths of
the exchanged particles the general structure is 
\beqn 
{\cal E}
&\sim&\quad Re({\cal C}) Re[\Delta(\alpha)\Delta(\beta)^{*}] Re({\cal
S}) - Re({\cal C}) Im[\Delta(\alpha)\Delta(\beta)^{*}] Im({\cal S})
\nonumber\\ &&+Im({\cal C})Re[\Delta(\alpha)\Delta(\beta)^{*}]
Im({\cal S}) -Im({\cal C}) Im[\Delta(\alpha)\Delta(\beta)^{*}]
Re({\cal S}), \la{eq_cp1b} \\
& \equiv & 
{\cal E}_{RRR} + {\cal
E}_{RII} + {\cal E}_{IRI} + {\cal E}_{IIR}.
\eeqn
If CP is conserved all
couplings can be chosen real \ci{Bartl_nucl} and ${\cal
E}_{IRI}={\cal E}_{IIR}=0$. The terms ${\cal E}_{RII}$ and ${\cal
E}_{IIR}$ are due to interference between s--channel exchange of
particles with finite width and the crossed channels. Their
contributions can be neglected except in the vicinity of the 
s-channel resonances (exchange of gauge bosons in the case
of $e^+ e^-$ or $q \bar{q}$ annihilation and 
in addition of Higgs bosons in $\mu^+ \mu^-$
annihilation), since they are proportional to the
width of the exchanged particles far from the pole. In 
${\cal E}_{RII}$ and ${\cal E}_{IRI}$ the imaginary part of
${\cal S}$ originates from
products of momenta and polarization vectors with the Levy--Civita
tensor $\epsilon_{\mu \nu \rho \sigma}$.  These two terms lead to
triple product correlations of momenta which are sensitive to CP
violation \ci{Christova}.
\subsubsection{CPT invariance}\la{sec:31}
Under CPT the helicity states of a Dirac fermion $f_D$ and a Majorana
fermion $f_M$, respectively, transform as
\beqn
&&|f_{D}(\vec{p},\lambda)> \stackrel{CPT}{\to}
(-1)^{\lambda-\frac{1}{2}}|\bar{f}_{D}(\vec{p},-\lambda)>, \la{hans_10}\\
&&|f_{M}(\vec{p},\lambda)> \stackrel{CPT}{\to}  
(-1)^{\lambda-\frac{1}{2}} \eta^{CPT}
|f_{M}(\vec{p},-\lambda)> \la{hans_1},
\eeqn
with the CPT phase
$\eta^{CPT}=\pm i$  
of Majorana fermions \ci{Kayser}.
Beyond that time reversal would interchange in-- and
outgoing states. However, if the finite width of the exchanged particles
is neglected CPT invariance and the unitarity of the S--matrix leads in leading
 order perturbation theory to symmetry relations for the helicity amplitudes
and for the production density matrix.

In the following we derive these symmetry relations from CPT
invariance for (charged or neutral) 
Dirac fermions and for the special case of Majorana fermions. \\[.2em]

\noindent {\bf \ref{sec:31}.1 Dirac fermions} 

\vspace{.5em}

\noindent CPT invariance relates in lowest order perturbation
theory the spin density matrix
$\rho_P^{\lambda_i\lambda_j,\lambda'_i\lambda'_j}$ of the process 
\bequ
f\bar{f}\to f_i(\vec{p}_i,\lambda_i) \bar{f}_j(\vec{p}_j,\lambda_j)
\la{eqa}
\eequ
to the spin density matrix
$\bar{\rho}_P^{\lambda_i\lambda_j,\lambda'_i\lambda'_j}$ of 
\bequ
f \bar{f}\to \bar{f}_i(\vec{p}_i,\lambda_i) f_j(\vec{p}_j,\lambda_j)
\la{eqb}
\eequ
by
\bequ 
\rho_{P}^{\lambda_i \lambda_j,
\lambda^{'}_i \lambda^{'}_j} (P_{f}^{1,2,3}; P_{\bar{f}}^{1,2,3};
\Theta)=
\bar{\rho}_{P}^{-\lambda_i -\lambda_j, -\lambda^{'}_i -\lambda^{'}_j}
(P_{\bar{f}}^{1,2},-P_{\bar{f}}^{3}; P_{f}^{1,2}, -P_{f}^{3};
\bar{\Theta})^{*}, \la{hans_7} 
\eequ 
where $\Theta$ denotes the angle
between $f$ and $f_i$ in (\ref{eqa}) and
$\bar{\Theta}\equiv\pi-\Theta$ is the angle between the direction of
beam particle $f$ and the outgoing antifermion $\bar{f}_i$ of the
conjugated process (\ref{eqb}) in the CMS.  For polarized beams the
polarizations of $f$ and $\bar{f}$ are interchanged with sign
reversal of the longitudinal polarization.  
To derive (\ref{hans_7}) the CPT transformation has been
supplemented by a rotation ${\cal R}_2(\pi)$ around the normal to the
production plane so that the beam direction is unchanged (see
Fig.~\ref{fig_1}a).

\begin{sloppypar}
Both density matrices $\rho_P$ and $\bar{\rho}_P$ can be expanded
according to (\ref{eq4_4h}) with the coefficients
$P(f_i\bar{f}_j)$, $\Sigma_P^a(f_i)$, $\Sigma_P^b(\bar{f}_j)$,
$\Sigma_P^{ab}(f_i\bar{f}_j)$ and 
$\bar{P}(\bar{f}_i f_j)$, $\bar{\Sigma}_P^a(\bar{f}_i)$, 
$\bar{\Sigma}_P^b(f_j)$,
$\bar{\Sigma}_P^{ab}(\bar{f}_if_j)$.
Then from CPT and ${\cal R}_2(\pi)$ invariance 
and with the substitutions
\bequ
\mbox{\hspace*{1cm}}(P_{f}^{1,2,3}; P_{\bar{f}}^{1,2,3}; \Theta)\to
(P_{\bar{f}}^{1,2},-P_{\bar{f}}^{3}; P_{f}^{1,2},-P_{f}^{3};
\bar{\Theta}=\pi-\Theta),\la{eq_fin}
\eequ
one obtains the following relations:
\beqn
&&P(f_i \bar{f}_j)=\bar{P}(\bar{f}_i f_j),\label{eq_dir1a}\\ 
&&\Sigma_P^a(f_i)=\eta_a\bar{\Sigma}_P^a(\bar{f}_i),\label{eq_dir1b}\\
&&\Sigma_P^b(\bar{f}_j)=\eta_b\bar{\Sigma}_P^b(f_j),\label{eq_dir1c}\\
&&\Sigma_P^{ab}(f_i \bar{f}_j)=\eta_a \eta_b\bar{\Sigma}_P^{ab}(\bar{f}_i f_j),
\label{eq_dir1d}
\eeqn with $\eta_1=1=\eta_2$ and $\eta_3=-1$.\\[.2em] 
\end{sloppypar}

\pagebreak

\noindent {\bf \ref{sec:31}.2 Majorana fermions} 

\vspace{.5em}

\noindent Since it is  
$\bar{\rho}_P\equiv \rho_P$  in (\ref{hans_7}) 
the invariance under CPT$\times{\cal R}_2(\pi)$ leads in lowest
order perturbation theory to constraints for the production density
matrix.  
The different terms in the expansion (\ref{eq4_4h})
of $\rho_P$ can be classified 
into contributions which are symmetric (Type S) and
antisymmetric (Type A) with regard to the substitution  
\bequ
\mbox{\hspace*{2cm}}(P_{f}^{1,2,3}; P_{\bar{f}}^{1,2,3}; \Theta)\to 
(P_{\bar{f}}^{1,2},-P_{\bar{f}}^{3}; P_{f}^{1,2},-P_{f}^{3}; \pi-\Theta),
\label{eq_sub1}
\eequ 
where $\Theta$ denotes the angle between $f$ and $f_i$ in the CMS.
\begin{itemize}
\item Type S: $P$, $\Sigma_P^1$, $\Sigma_P^2$, $\Sigma_P^{11}$,
$\Sigma_P^{22}$, $\Sigma_P^{33}$, $\Sigma_P^{12}$, $\Sigma_P^{21}$ 
\item Type A: $\Sigma_P^3$, $\Sigma_P^{13}$, $\Sigma_P^{31}$, 
$\Sigma_P^{23}$, $\Sigma_P^{32}$ 
\end{itemize}
All of them include terms ${\cal E}_{RRR}$ and 
${\cal E}_{IRI}$ of (\ref{eq_cp1b}).

For unpolarized beams the terms of type S are forward--backward 
symmetric whereas those
of type A are forward--backward antisymmetric.  
These symmetry properties
hold also for production of Majorana fermions with 
polarized $e^+ e^-$ beams. 

The forward--backward symmetry of the term $P$ results in the FB--symmetry
of the differential cross sections for the production of Majorana fermions with
unpolarized and longitudinally
polarized beams \ci{Christova}. Beyond that, also
the symmetry properties
of the different components of their polarization and of the spin--spin 
correlations are specific for their Majorana character.

\subsubsection{CP invariance}\la{sec:32}
We now study the consequences of the invariance under a CP transformation
(illustrated in Fig.~\ref{fig_1}b) for the production of two Dirac
or two Majorana fermions.

Under CP the helicity states of a Dirac and a Majorana fermion transform as
\beqn
&&|f_{D}(\vec{p},\lambda)> \stackrel{CP}{\to}
|\bar{f}_{D}(-\vec{p},-\lambda)>,\la{hans_6}\\
&&|f_{M}(\vec{p},\lambda)> \stackrel{CP}{\to}\eta^{CP} 
|f_{M}(-\vec{p},-\lambda)> \la{nachtrag_1}
\eeqn
with $\eta^{CP}=\pm i$.\\[.2em]
\newpage
\noindent {\bf \ref{sec:32}.1 Dirac fermions}

\vspace{.5em}

\noindent CP invariance results in a relation
between the spin density matrix
$\rho_P$ for the conjugated process
\bequ
f \bar{f} \to f_i(\vec{p}_i,\lambda_i) \bar{f}_j(\vec{p}_j,\lambda_j)
\eequ
and the density matrix 
$\bar{\rho}_P$ for 
\bequ
f \bar{f} \to \bar{f}_i(\vec{p}_i,\lambda_i) f_j(\vec{p}_j, \lambda_j)
\eequ
\bequ 
\rho_{P}^{\lambda_i \lambda_j, \lambda^{'}_i \lambda^{'}_j}
(P_{f}^{1,2,3}; P_{\bar{f}}^{1,2,3}; \Theta)=
\bar{\rho}_{P}^{-\lambda_i -\lambda_j, -\lambda^{'}_i -\lambda^{'}_j}
(P_{\bar{f}}^1,-P_{\bar{f}}^{2,3};P_{f}^1,-P_{f}^{2,3}; \bar{\Theta}),
\la{hans_8}
\eequ
with the same denotation as in (\re{hans_7}). For polarized beams the 
polarizations of $f$ and $\bar{f}$ are interchanged with an additional
sign reversal for the longitudinal polarization and the polarization
perpendicular to the production plane.

Expanding both density matrices $\rho_P$ and $\bar{\rho}_P$ 
according to (\ref{eq4_4h}) one obtains 
with the same substitutions 
\bequ
\mbox{\hspace*{1cm}}(P_{f}^{1,2,3}; P_{\bar{f}}^{1,2,3};
\Theta)\to(P_{\bar{f}}^1,-P_{\bar{f}}^{2,3};P_{f}^1,-P_{f}^{2,3}; 
\bar{\Theta}=\pi-\Theta)
\la{eq_fin2}
\eequ
as in (\ref{hans_8})
the following relations between the 
coefficients:
\beqn
&&P(f_i \bar{f}_j)=\bar{P}(\bar{f}_i f_j),\label{eq_dir2a}\\
&&\Sigma_P^{a}(f_i )=\eta_a \bar{\Sigma}_P^{a}(\bar{f}_i ),\label{eq_dir2b}\\
&&\Sigma_P^{b}(\bar{f}_j )=\eta_b \bar{\Sigma}_P^{b}(f_j ),\label{eq_dir2c}\\
&&\Sigma_P^{ab}(f_i \bar{f}_j)=\eta_a \eta_b\bar{\Sigma}_P^{ab}(\bar{f}_if_j), 
\label{eq_dir2d}
\eeqn
with $\eta_1=+1$ and $\eta_2=-1=\eta_3$.
For unpolarized or longitudinally polarized $e^+ e^-$ 
beams, the dependence on the beam
polarization is given by the two factors $(1-P^3_{f} P^3_{\bar{f}})$ and
$(P^3_{f}-P^3_{\bar{f}})$. If the widths of the
exchanged particles in the s--channel are neglected, 
the CPT symmetry
relations (\re{hans_7}) hold and one obtains
\bequ
\mbox{\hspace*{1cm}}\Sigma_P^2=0,\quad \Sigma_P^{12}=0=\Sigma_P^{21},\quad 
\Sigma_P^{23}=0=\Sigma_P^{32}.
\la{neu_1}
\eequ

\vspace{1em}

\noindent {\bf \ref{sec:32}.2 Majorana fermions}

\vspace{.5em}

With $\bar{\rho}_P\equiv \rho_P$ in (\ref{hans_8}) 
CP invariance again leads 
to a classification of the different contributions in the expansion
(\ref{eq4_4h}) of $\rho_P$ according to their CP symmetry properties
with regard to the substitution 
\bequ
\mbox{\hspace*{1cm}}(P_{f}^{1,2,3}; P_{\bar{f}}^{1,2,3}; \Theta)\to
(P_{\bar{f}}^1,-P_{\bar{f}}^{2,3}; P_{f}^1,-P_{f}^{2,3}; \pi-\Theta),
\label{sub_2}
\eequ
where $\Theta$ is the scattering angle 
in the CMS:
\begin{itemize}
\item Type S: $P$, $\Sigma_P^1$, $\Sigma_P^{11}$,
$\Sigma_P^{22}$, $\Sigma_P^{33}$, $\Sigma_P^{23}$, $\Sigma_P^{32}$, 
\item Type A: $\Sigma_P^2$, $\Sigma_P^3$, $\Sigma_P^{12}$, $\Sigma_P^{21}$,
$\Sigma_P^{13}$, $\Sigma_P^{31}$
\end{itemize}
All terms have contributions with the structure 
${\cal E}_{RRR}$ and ${\cal E}_{RII}$ of (\ref{eq_cp1b}).

\begin{sloppypar}
For unpolarized and longitudinally polarized $e^+e^-$ beams, the terms
of type~S are FB--symmetric and those of type A are
FB--antisymmetric. For both Dirac and Majorana fermions 
the terms $\Sigma^2_P$,
$\Sigma^{12}_P$, $\Sigma^{21}_P$, $\Sigma^{23}_P$ and $\Sigma^{32}_P$ 
vanish for
unpolarized or longitudinally polarized beams if the CPT relations 
(\ref{eq_dir1a})--(\ref{eq_dir1d}) hold.  The FB--symmetry or antisymmetry
of the polarization and of the spin--spin correlations, however, is
specific for the Majorana character of the produced fermions beyond
the FB--symmetry of their production cross section. Since in general
the production of Dirac fermions will not exhibit these symmetry
properties the measurement of polarization and spin--spin correlations
are helpful for an experimental verification of the Majorana character.
\end{sloppypar}

\subsubsection{CP violation}\la{sec:321}
If CP is violated due to complex couplings 
the two types ${\cal E}_{IRI}$, ${\cal E}_{IIR}$
of CP violating terms contribute in (\re{eq_cp1b}), 
whereas far from
the s-channel resonances 
${\cal E}_{IIR}$ 
can be neglected as explained in section \ref{sec:30}.

%
%
\vspace{.5em}

Since for Dirac fermions the CP transformation relates the two processes 
$f \bar{f}\to f_i \bar{f}_j$ and $f \bar{f}\to \bar{f}_i f_j$
we restrict ourselves to the discussion of
 the consequences of CP violation for the production 
density matrix of {\it Majorana fermions} for unpolarized or longitudinally
polarized beams. 
Both types of CP violating terms ${\cal E}_{IRI}$ and ${\cal E}_{IIR}$ lead
to contributions which
show different FB angular dependence compared to the CP conserving ones.

In particular terms with the structure ${\cal E}_{IRI}$ 
result in nonvanishing polarization 
$\Sigma^2_P$ perpendicular to the production plane and nonvanishing 
spin--spin correlations $\Sigma_P^{12,21}$ and $\Sigma_P^{23,32}$:
\begin{itemize}
\item FB--symmetric in ${\cal E}_{IRI}$: 
$\Sigma^2_P$, $\Sigma^{12}_P$, $\Sigma^{21}_P$ 
\item FB--antisymmetric in  ${\cal E}_{IRI}$:
$\Sigma^{23}_P$, $\Sigma^{32}_P$ 
\end{itemize}
These terms are proportional
to $Im({\cal S})$ and are sensitive to triple product correlations
of momenta. They are CP--odd quantities \ci{CKMZ}.

All other contributions in (\ref{eq4_4h}) are CP--even quantities and
get in the case of CP violation
terms with the structure ${\cal E}_{IIR}$:
\begin{itemize}
\item FB--symmetric in ${\cal E}_{IIR}$:
$\Sigma^3_P$, $\Sigma^{13}_P$, $\Sigma^{31}_P$ 
\item FB--antisymmetric in ${\cal E}_{IIR}$:
$P$, $\Sigma^1_P$, $\Sigma^{11}_P$, $\Sigma^{22}_P$,
$\Sigma^{33}_P$ 
\end{itemize} 
The consequences of 
CP and CPT for the FB--symmetry of the different terms of
the production spin density matrix of
Majorana fermions are summarized in 
Table~\re{tab_-1}. 
\subsection{CPT and CP symmetries of the decay matrices}
\la{sec:33}
In this section we derive from CPT and CP invariance constraints for the decay
matrices for the three--body decay 
of neutral and charged Dirac fermions and of Majorana fermions. 
\subsubsection{CPT invariance}
\la{sec:331}


\noindent {\bf \ref{sec:331}.1 Dirac fermions}

\vspace{.5em}

\noindent For the three--body decay
\bequ
f_i(\vec{p}_i,\lambda_i)\to f_{i1}(\vec{p}_{i1})f_{i2}(\vec{p}_{i2})
\bar{f}_{i3}(\vec{p}_{i3})\la{eq_neu2}
\eequ
of a Dirac fermion $f_i$ into a Dirac fermion $f_{i1}$ and a
fermion--antifermion pair $f_{i2}\bar{f}_{i3}$, CPT invariance
relates the decay matrix $\rho_D$ of (\ref{eq_neu2}) 
with the decay matrix 
$\bar{\rho}_D$ of the process
\bequ
\bar{f}_i(\vec{p}_i,\lambda_i)\to 
\bar{f}_{i1}(\vec{p}_{i1})\bar{f}_{i2}(\vec{p}_{i2})f_{i3}(\vec{p}_{i3}).
\la{eq_neu3}
\eequ
Since we do not study the polarization of the decay fermions
we sum about their helicities and obtain 
\bequ
\rho_{D,\lambda_i^{'}\lambda_i}(\vec{p}_{i2},\vec{p}_{i3})=
(-1)^{\lambda_i-\lambda_i^{'}}
\bar{\rho}_{D,-\lambda_i^{'}-\lambda_i}^{*}(\vec{p}_{i3},\vec{p}_{i2}), 
\la{zerfall_1a}
\eequ
with the momenta of the fermion--antifermion pair interchanged and the signs
of the helicities reversed.

Expanding $\rho_D$ and $\bar{\rho}_D$ according to (\re{eq4_4i}) leads for the
coefficients to
\beqn
D(\vec{p}_{i2},\vec{p}_{i3}) & = & \bar{D}(\vec{p}_{i3},\vec{p}_{i2}),
\label{eq_dir4a}\\
\Sigma_D^{a}(\vec{p}_{i2},\vec{p}_{i3}) & = & -\bar{\Sigma}_D^a
(\vec{p}_{i3},\vec{p}_{i2}), \quad a=1,2,3.\label{eq_dir4c} 
\eeqn 

\vspace{1em}
 
\noindent {\bf \ref{sec:331}.2 Majorana fermions}

\vspace{.5em}

\noindent For the decay 
\bequ
f_i(\vec{p},\lambda_i)\to f_{i1}(\vec{p}_{i1})f_{i2}(\vec{p}_{i2})
\bar{f}_{i2}(\vec{p}_{i3}) \label{eq_neu}
\eequ 
of a
Majorana fermion $f_i$ into a Majorana fermion $f_{i1}$ and a Dirac
fermion--antifermion pair $f_{i2}\bar{f}_{i2}$ with $\bar{\rho}_D\equiv \rho_D$
in (\ref{zerfall_1a}),
CPT invariance leads to constraints for the decay
matrix from which one obtains definite
symmetry properties for the expansion coefficients of $\rho_D$ in 
(\ref{eq4_4i})
\beqn
D(\vec{p}_{i2},\vec{p}_{i3}) & = & D(\vec{p}_{i3},\vec{p}_{i2}) \\ 
\Sigma_D^{a}(\vec{p}_{i2},\vec{p}_{i3}) & = & 
-\Sigma_D^{a}(\vec{p}_{i3},\vec{p}_{i2}),\quad a=1,2,3
\eeqn
if the momenta of the fermion--antifermion pair $f_{i2}\bar{f}_{i2}$
are exchanged.

\subsubsection{CP invariance}
\la{sec:332}
Applying the space--inversion $P$ to the decay (\ref{eq_neu2})
also the momentum of the decaying particle $f_i$ is reversed. We
therefore add a rotation ${\cal R}_{\vec{n}}(\pi)$ around the normal
to the plane $(\vec{p}_i,\vec{p}_{i1})$. Then 
${\cal P}\times {\cal R}_{\vec{n}}(\pi)$ is equivalent to a reflection
at the plane $(\vec{p}_i,\vec{p}_{i1})$.

\vspace{1em}

\noindent {\bf \ref{sec:332}.1 Dirac fermions}

\vspace{.5em}
 
\noindent The combined transformation
$CP\times {\cal R}_{\vec{n}}(\pi)$ leads 
for the decay of a Dirac fermion to the process
\bequ
\bar{f}_i(\vec{p}_i,-\lambda_i)\to \bar{f}_{i1}(\vec{p}_{i1})
\bar{f}_{i2}(\hat{p}_{i2})f_{i3}(\hat{p}_{i3}),
\label{eq_neu_dir}
\eequ where $\hat{p}_{i2}$ and $\hat{p}_{i3}$ denotes the momentum
$\vec{p}_{i2}$ and $\vec{p}_{i3}$, respectively, reflected at the
plane $(\vec{p}_i,\vec{p}_{i1})$.  The invariance under $CP\times
{\cal R}_{\vec{n}}(\pi)$ relates the decay matrix $\rho_D$ for
(\ref{eq_neu2}) with the decay matrix $\bar{\rho}_D$ for the
antiparticle in (\ref{eq_neu_dir}) 
\bequ
\rho_{D,\lambda_i^{'}\lambda_i}(\vec{p}_{i2},\vec{p}_{i3})=
(-1)^{\lambda_i-\lambda_i^{'}}
\bar{\rho}_{D,-\lambda_i^{'}-\lambda_i}(\hat{p}_{i3},\hat{p}_{i2}). 
\la{eq_neu5}
\eequ 
In contrast to the CPT relation (\ref{zerfall_1a}) the
momenta of the fermion--antifermion pair are interchanged and
additionally reflected at the plane ($\vec{p}_i$, $\vec{p}_{i1}$).  For the
coefficients in the expansion (\ref{eq4_4i}) one obtains 
\beqn
D(\vec{p}_{i2},\vec{p}_{i3}) & = & \bar{D}(\hat{p}_{i3},\hat{p}_{i2}),
\label{eq_dir3a}\\ 
\Sigma_D^{2}(\vec{p}_{i2},\vec{p}_{i3}) & = & \bar{\Sigma}_D^{2}(\hat{p}_{i3},\
\hat{p}_{i2}),\label{eq_dir3b}\\  
\Sigma_D^{1,3}(\vec{p}_{i2},\vec{p}_{i3}) & = & -\bar{\Sigma}_D^{1,3}
(\hat{p}_{i3},\hat{p}_{i2}).\label{eq_dir3c}
\eeqn
Combining these CP relations with the CPT properties 
(\ref{eq_dir4a})--(\ref{eq_dir4c}) of the expansion
coefficients leads to 
\begin{itemize}
\item symmetric $D$, $\Sigma_D^{1,3}$ 
\item antisymmetric $\Sigma_D^2$ 
\end{itemize}
under the reflection of the momenta $\vec{p}_{i2}$ and $\vec{p}_{i3}$
at the plane $(\vec{p}_i,\vec{p}_{i1})$.

\vspace{1em}
 
\noindent {\bf \ref{sec:332}.2 Majorana fermions}

\vspace{.5em}

\noindent For the decay (\ref{eq_neu}) of a Majorana
fermion with $\bar{\rho}_D\equiv \rho_D$ in (\ref{eq_neu5})
the combined transformation $CP\times {\cal R}_{\vec{n}}(\pi)$
constraints the decay matrix and
their expansion coefficients 
when the momenta of the Dirac fermion $f_{i2}$ and antifermion
$\bar{f}_{i2}$ are interchanged and reflected at the plane of the Majorana 
fermions $f_i, f_{i1}$. 
Neglecting the width of the exchanged particles CP and CPT invariance
lead to the same symmetry properties of $D$, $\Sigma_D^{1,2,3}$
as in the case of Dirac fermions under the reflection of the 
momenta $\vec{p}_{i2}$, $\vec{p}_{i3}$
at the plane $(\vec{p}_i,\vec{p}_{i1})$.
\\[.2em]

\section{Energy and angular distributions of the decay products}
\la{sec:40}
In this section we study for the production of Dirac and Majorana fermions
and their subsequent
three--particle decay into fermions
the influence of the polarization and of the 
spin--spin correlations on 
the energy spectrum and angular distributions in the
lab frame.
If the decay of only 
one of the produced fermions, e.g. $f_i$, is considered, it is 
in (\re{eq4_5}) $\Sigma^b_D(f_j)\equiv 0$ and $D(f_j)\equiv 1$.
Then the total cross section for the combined process of
production and decay is given by 
$\sigma_P(f \bar{f}\to f_i f_j)\times BR(f_i\to f_{i1} f_{i2} f_{i3})$.
\subsection{Energy spectrum and opening angle distribution of the decay 
leptons}\la{sec:41}
We expect that in general the energy distribution of the decay
particles in the lab system depends on the polarization of the
decaying particle.  In its rest frame the decay
angular distribution is determined by the polarization with respect to
a suitably chosen quantization axis. Since boosting in this direction
into the lab frame the energy of a decay particle depends on the
orientation of its momentum with respect to this axis it also depends
on the polarization of the decaying particle. The same argument applies to 
the distribution
of the opening angle in the lab frame between the decay products.

In the following we show that due to the specific
CPT/CP properties of Majorana fermions
the energy distribution and the opening angle distribution
of the decay fermions are independent  of the polarizations of the decaying
particle. In this case both distributions factorize into a production 
and a decay piece. 

For our analysis we parametrize the phase space in the lab frame in
a way that it
factorizes into the production and decay phase space.

\subsubsection{Energy spectrum}
\la{sec:411}

\begin{sloppypar}
For the energy distribution $d\sigma/dE_{i2}$
the phase space can be parametrized
by the polar angle $\Theta_{i,i2}$  ($\Theta_{i,i3}$)  between the momenta of 
$f_i$ and $f_{i2}$ ($f_{i3}$) and the azimuth $\Phi_{i,i2}$ ($\Phi_{i,i3}$)
of $f_{i2}$ ($f_{i3}$). All polar (azimuthal) angles are 
denoted by $\Theta_{\alpha \beta}$ ($\Phi_{\alpha \beta}$), 
where the first index denotes the polar axis:
\bequ
d\sigma= {\cal F} |T|^2 
 \sin\Theta d\Theta d\Phi \sin\Theta_{i,i2} d\Theta_{i,i2}d\Phi_{i,i2}
 \sin\Theta_{i,i3}d\Theta_{i,i3}d\Phi_{i,i3}dE_{i2},
\la{eq23_1}
\eequ
where $\Theta$ is the production angle. 
For light fermions $f_{i2}$ and $f_{i3}$ 
${\cal F}$ is 
given by: 
\bequ
{\cal F}=\frac{q}{2^{11} (2\pi)^7  m_i \Gamma_{i}E_b^3}
\frac{E_{i2} [m_i^2-m_{i1}^2-2 E_{i2}(E_i-q\cos\Theta_{i,i2})]}
{[E_i-q\cos\Theta_{i,i3}-E_{i2}(1-\cos\Theta_{i2,i3})]^2}. \la{eq_F}
\eequ
$E_b$ denotes the beam energy,
$E_{\alpha}$ denotes the energy of the fermion $f_{\alpha}$ in the lab frame
and $q=|\vec{p}_i|=|\vec{p}_j|$. 
The opening angle $\Theta_{i2,i3}$ between $f_{i2}$ and $f_{i3}$ can be
expressed via (\ref{eq23_3}) by the independent variables chosen in 
(\ref{eq23_1}).

The kinematical boundaries for $\Theta_{i,i2}$ depend on $E_{i2}$:
{\small \begin{equation}
\Theta_{i,i2}=\left\{ \begin{array}{l@{\quad}l}
[0,\pi],& {\scriptstyle 0<E_{i2}<\frac{m_i^2-m_{i1}^2}{2 (E_i+q)}}\\
(0,\arccos(\frac{E_i}{q}-\frac{m_i^2-m_{i1}^2}{2 q E_{i2}})), &
{\scriptstyle \frac{m_i^2-m_{i1}^2}{2(E_i+q)}\le E_{i2}\le
\frac{m_i^2-m_{i1}^2}{2(E_i-q)}}. \end{array}\right .\la{eq23_2}
\end{equation}}
\end{sloppypar}
\subsubsection{Opening angle distribution}

For the distribution of the opening angle $\Theta_{i2,i3}$ between the 
fermions $f_{i2}$ and $f_{i3}$ from the decay 
$f_i\to f_{i1} f_{i2} f_{i3}$ it is favorable to parametrize the phase 
space of $f_{i3}$ by
$\Theta_{i2,i3}$ 
and the corresponding 
azimuthal angle $\Phi_{i2,i3}$ (Fig.~\re{fig_2}), i.e. to substitute in
(\ref{eq23_1}) $\Theta_{i,i3}\to \Theta_{i2,i3}$ and 
$\Phi_{i,i3}\to\Phi_{i2,i3}$.
The kinematical region of $E_{i2}$ depends on $\Theta_{i,i2}$:
$0\le E_{i2} \le \frac{m_i^2-m_{i1}^2}{2 (E_i-q\cos\Theta_{i,i2})}$. 
The angles 
$\Theta_{i,i3}$ and $\Phi_{i,i3}$ can be expressed by  
(\ref{eq322_3a}) and (\ref{eq322_3b}), (\ref{eq322_3c})--(\ref{eq322_44}) 
by the chosen independent angular variables.\\

\subsubsection{Transverse polarizations} 

In  the contributions $\Sigma^a_P(f_i)\Sigma_D^a(f_i)$, $a=1,2,3$, 
from the polarization of the decaying particle $f_i$ both
factors depend on the polarization vectors $s^{a\mu}(f_i)$ defined with
respect to the production plane, cf. section~2.1.
They can be written
\bequ
\Sigma_P^a(f_i)=\tilde{\Sigma}_P^{\mu}(f_i)s^a_{\mu}(f_i),\quad\mbox{and}\quad
\Sigma_D^a(f_i)=\tilde{\Sigma}_D^{\mu}s^a_{\mu}(f_i).
\label{eq_neu6}
\eequ
To study their influence on the energy and the opening angle distribution we
distinguish between the contributions $\Sigma_P^{1,2}(f_i)$ from the
transverse polarizations and $\Sigma^3_P(f_i)$ from the longitudinal 
polarization.
In the contributions of the transverse polarizations,
the coefficients $\Sigma^{1,2}_D(f_i)$
depend on the azimuth $\Phi_{i,i2}$ between the production plane and the 
plane defined by the decaying fermion $f_i$ and the decay product $f_{i2}$.
To separate the $\Phi_{i,i2}$ dependence we introduce a new system of 
polarization vectors $t^{a \nu}(f_i)$ with
$t^{3 \nu}(f_i)=s^{3 \nu}(f_i)$, whereas in the lab system $\vec{t^2}(f_i)$ is 
perpendicular to the plane defined by the momenta of $f_{i}$ and 
$f_{i2}$ and $\vec{t^1}(f_1)$ is in this plane orthogonal to the 
momentum of the decaying fermion $f_i$. Then in (\re{eq_neu6}) the 
transverse polarization vectors $s^{1,2 \nu}(f_i)$ are
\beqn
s^{1\nu}(f_i)&=&
\cos\Phi_{i,i2} t^{1\nu}(f_i)-\sin\Phi_{i,i2} t^{2\nu}(f_i),
\la{eq3_1c}\\
s^{2\nu}(f_i)&=&
\sin\Phi_{i,i2} t^{1\nu}(f_i)+\cos\Phi_{i,i2} t^{2\nu}(f_i).
\la{eq3_1d}
\eeqn
Applying the transformations (\ref{eq23_3}) and
(\ref{eq322_3a}), (\ref{eq322_3b}), (\ref{eq322_3c}) -- (\ref{eq322_44}) 
for the angles $\Theta_{i,i3}$ and $\Phi_{i,i3}$
to the phase space parameterization
(\ref{eq23_1}) the contributions of 
$\Sigma^{1,2}_D(f_i)$ to the opening angle distribution and to the lepton 
energy spectrum vanish for both Majorana and Dirac fermions
due to the integration over $\Phi_{i,i2}$ \ci{Gudi_diss}.

\subsubsection{Longitudinal polarization}
\la{sec:414}

\noindent {\bf \ref{sec:414}.1 Majorana fermions}

\vspace{.5em}

\noindent The longitudinal polarization $\Sigma^3_P(f_i)$ is 
forward--backward antisymmetric if CP is conserved, cf. section~\ref{sec:32}. 
Due to the 
factorization of the phase space in production and decay
also the contribution of the 
longitudinal polarization vanishes after integration over 
the production angle $\Theta$.

{\it Consequently both the
energy and the opening angle distribution of the decay products
of Majorana fermions in the laboratory system 
are independent of spin correlations and factorize exactly
into production and decay if CP is conserved.} 

The factorization of the energy and opening angle distribution is
essentially equivalent since for both observables it is a
consequence of the FB-antisymmetry of the longitudinal polarization of
Majorana fermions.

\begin{sloppypar}
Assuming CP violation and taking into account the widths
of the exchanged particles 
the exact factorization of the energy distribution and of the  
opening angle distribution of the decay products 
of Majorana fermions is destroyed,
since the longitudinal
polarization $\Sigma^3_P$ gets FB--symmetric
contributions from CP violating terms. 
These additional terms have the structure ${\cal E}_{IIR}$,
and are therefore for energies far from the resonance
proportional to the width of the exchanged particle.
\end{sloppypar}

\vspace{1em}
 
\noindent {\bf \ref{sec:414}.2 Dirac fermions}

\vspace{.5em}

\noindent In general, the longitudinal polarization  
of produced Dirac fermions
is not forward--backward asymmetric. Therefore the 
energy spectra and opening angle distributions of the decay products 
do not factorize into production and decay. 

\subsection{Decay lepton angular distribution}
\la{sec:4}
For the distribution $d\sigma/d\cos\Theta_{1,i2}$ of the angle
$\Theta_{1,i2}$ in the lab frame between the incoming fermion $f$
and the fermion $f_{i2}$ from the decay 
$f_i\to f_{i1} f_{i2} f_{i3}$ we parameterize the phase space by
\bequ
d\sigma = {\cal F} |T|^2 
\sin\Theta d\Theta d\Phi \sin\Theta_{1,i2} d\Theta_{1,i2}d\Phi_{1,i2}
 \sin\Theta_{1,i3}d\Theta_{1,i3}d\Phi_{1,i3}dE_{i2}.\la{eq23_3a}
\eequ
and express all angles in ${\cal F}$ (\re{eq_F}) 
and the azimuth $\Phi_{i,i2}$ in (\re{eq3_1c}), (\re{eq3_1d}) by
the independent variables in (\ref{eq23_3a}).
The relations are given in (\ref{eq32_3b}), (\ref{eq322_3aa}), 
(\ref{eq322_3d}), (\ref{eq322_3e}) \ci{Gudi_diss}.
Thus the kinematic factor ${\cal F}$ depends explicitly on the scattering 
angle $\Theta$ and
neither the contributions $\Sigma_D^{1,2}(f_i)$ of
the transverse polarizations
nor that of the longitudinal polarization $\Sigma_P^3(f_i)$ vanish 
due to phase space integration.\\[.2em]
{\it Consequently neither for Dirac fermions  
nor for Majorana fermions the decay
lepton angular distribution in the lab frame factorizes in production
and decay but depends sensitively on the spin correlations.}
\subsection{Siamese opening angle distribution}
\la{sec:43}
The siamese opening angle  $\Theta_{j2,i2}$
denotes the angle between decay products $f_{i2}$ and
$f_{j2}$ from the decay of different particles $f_i$ and 
$\bar{f}_j$ ($f_j$ in the case of Majorana fermions) in the lab frame.
The distribution $d\sigma/d\cos\Theta_{j2,i2}$ is determined by the spin--spin
correlations $\Sigma_P^{ab}$ (\ref{eq4_5})
between the decaying mother particles $f_i$ and $\bar{f}_j$.
Here it is favorable to parameterize both
phase spaces for 
the decay of $f_i$  and of $\bar{f}_j$ 
by the angles $\Theta_{i,i3}$, $\Phi_{i,i3}$, 
and $\Theta_{i,j2}$, $\Phi_{i,j2}$ 
\beqn
d\sigma&=& {\cal F}{\cal G}|T|^2 
\sin\Theta d\Theta d\Phi\sin\Theta_{i, j2} d\Theta_{i,j2}d\Phi_{i,j2}
 \sin\Theta_{i,j3}d\Theta_{i,j3}d\Phi_{i,j3}dE_{j2}\nonumber\\
&&\times \sin\Theta_{i,i3} d\Theta_{i,i3}d\Phi_{i,i3}
\sin\Theta_{j2,i2} d\Theta_{j2,i2}d\Phi_{j2,i2} dE_{i2},
\la{eq34_5}
\eeqn
where 
$\Theta_{i,j3}$, $\Phi_{i,j3}$
is defined with respect to the direction of $f_i$ and $\Theta_{j2,i2}$,
$\Phi_{j2,i2}$.
${\cal F}$ is given by (\re{eq_F}) and 
\beqn
{\cal G}&=&\frac{1}{2^5 (2 \pi)^5 m_j \Gamma_j}
\frac{E_{j2} [m_j^2-m_{j1}^2-2 E_{j2}(E_j-q\cos\Theta_{i,j2})]}
{[E_j-q\cos\Theta_{i,j3}-E_{j2}(1-\cos\Theta_{j2,j3})]^2}. \la{eq_G}
\eeqn
Using the transformation 
formulae (\ref{eq23_3}) and 
(\ref{eq325_3b}) -- (\ref{eq325_3d}) all angles can be expressed by 
the independent variables chosen in (\ref{eq34_5}).
 
With the same arguments as for the contributions from transverse 
polarization to the opening angle distribution one infers:\\[.2em]
{\it The siamese opening angle distribution 
factorizes neither for Majorana nor for Dirac 
fermions, it depends, however, only 
on the diagonal spin--spin correlations $\Sigma_P^{11}$, 
$\Sigma_P^{22}$, $\Sigma_P^{33}$ if CP is conserved.} 

\vspace{.2cm} The polarizations of the produced fermions and the
spin--spin correlations between these contributing to the energy and
the different angular distributions of their decay products are listed
in Table~\re{tab_1}.

\section{Conclusion}
\la{sec:7}
Assuming CPT and CP invariance we derived symmetry properties
of the spin density matrix
for production of Dirac and Majorana
fermions in fermion--antifermion annihilation with polarized beams
and for their three--particle decay matrices.
Majorana fermions show specific forward--backward symmetry
properties of their polarizations and the spin--spin correlations.
In particular for the
production of Majorana fermions with unpolarized or longitudinally
polarized $e^+e^-$ beams their longitudinal polarization in the CMS is
forward-backward antisymmetric 
so that 
the energy distributions of the decay products and the
distribution of the opening angle between them are exactly independent of
spin correlations if CP is conserved and if the width of the exchanged
particles can be neglected. Thus they factorize into a production and
a decay piece which allows to study the dynamics of the decay of Majorana
fermions independently of that of the production process. 
Since this
factorization is specific for Majorana fermions it opens the possibility
to establish experimentally the Majorana character by comparing the
measured decay energy and opening angle distributions with Monte Carlo
studies.

\section{Acknowledgment}
\label{ack}
We thank A.~Bartl, E.~Christova, F. Franke 
and W.~Majerotto for valuable discussions.
This work was supported by the Bundesministerium f\"ur Bildung
und Forschung, contract No.\ 05 7WZ91P (0), by DFG FR 1064/4--1, by
the Fonds zur F\"orderung der wissenschaftlichen Forschung of Austria,
project No.\ P13139-PHY and by the EU TMR Network Contract No.
HPRN-CT-2000-00149.

\begin{appendix}
\setcounter{equation}{0}
\renewcommand{\theequation}{\thesection.\arabic{equation}}
\section{Parametrization of phase space}\la{anh_A}
For completeness we list all angular relations used 
for the different parametrizations of the phase space in section~3.  
\beqn
\cos\Theta_{i2,i3}&=&\cos\Theta_{i,i2}\cos\Theta_{i,i3}+
\sin\Theta_{i,i2}\sin\Theta_{i,i3}\cos(\Phi_{i,i2}-\Phi_{i,i3}),
\la{eq23_3}\\
&=&\cos\Theta_{1,i2}\cos\Theta_{1,i3}
+\sin\Theta_{1,i2}\sin\Theta_{1,i3}\cos(\Phi_{1,i2}-\Phi_{1,i3})
 \la{eq32_3b},\\
\cos\Theta_{i,i3}&=&\cos\Theta_{i,i2}\cos\Theta_{i2,i3}+
\sin\Theta_{i,i2}\sin\Theta_{i2,i3}\cos\Phi_{i2,i3},\la{eq322_3a}\\
&=&\cos\Theta\cos\Theta_{1,i3}+
\sin\Theta\sin\Theta_{1,i3}\cos\Phi_{1,i3},\la{eq322_3aa}\\
\sin\Phi_{i,i3}&=&\frac{\sin\Theta_{1,i3}}{\sin\Theta_{i,i3}}\sin\Phi_{1,i3},
\la{eq322_3d}\\
&=&\sin\Phi_{i,i2}\cos(\Phi_{i,i2}-\Phi_{i,i3})-
\cos\Phi_{i,i2}\sin(\Phi_{i,i2}-\Phi_{i,i3}),\la{eq322_3b}\\
\cos\Phi_{i,i3}&=&-[\cos\Theta_{1,i3}\sin\Theta-\sin\Theta_{1,i3}\cos\Theta
\cos\Phi_{1,i3}]/\sin\Theta_{i,i3},\la{eq322_3e}\\
&=&\cos\Phi_{i,i2}\cos(\Phi_{i,i2}-\Phi_{i,i3})+
\sin\Phi_{i,i2}\sin(\Phi_{i,i2}-\Phi_{i,i3}).\la{eq322_3c}
\end{eqnarray}
In (\ref{eq322_3b}) and (\ref{eq322_3c}) one has to insert:
\begin{eqnarray}
\cos(\Phi_{i,i2}-\Phi_{i,i3})&=& \frac{\cos\Theta_{i2,i3}\sin\Theta_{i,i2}
-\cos\Theta_{i,i2}\sin\Theta_{i2,i3}
\cos\Phi_{i2,i3}}{\sin\Theta_{i,i3}},\la{eq322_4}\\
\sin(\Phi_{i,i2}-\Phi_{i,i3})&=&\frac{\sin\Theta_{i2,i3}}{\sin\Theta_{i,i3}}
\sin\Phi_{i2,i3}\la{eq322_44}.
\end{eqnarray}
The angles $\Theta_{i,i2}$, $\Phi_{i,i2}$ 
are given analogously to (\ref{eq322_3aa}), (\ref{eq322_3d}) and 
(\ref{eq322_3e}) with $\Theta_{1,i3}\to \Theta_{1,i2}$,
$\Phi_{1,i3}\to \Phi_{1,i2}$ and $\Theta_{i,i3}\to \Theta_{i,i2}$.
\begin{eqnarray}
\cos\Theta_{i,i2}&=&\cos\Theta_{i,j2}\cos\Theta_{j2,i2}
+\sin\Theta_{i,j2}\sin\Theta_{j2,i2}\cos\Phi_{j2,i2},\la{eq325_3b}\\
\sin\Phi_{i,i2}&=&\mbox{\hspace{-.4cm}}-\sin(\Phi_{i,j2}-\Phi_{i,i2})\cos\Phi_{i,j2}
+\cos(\Phi_{i,j2}-\Phi_{i,i2})\sin\Phi_{i,j2},\la{eq325_3e}\\
\cos\Phi_{i,i2}&=&\sin(\Phi_{i,j2}-\Phi_{i,i2})\sin\Phi_{i,j2}
+\cos(\Phi_{i,j2}-\Phi_{i,i2})\cos\Phi_{i,j2},\la{eq325_3f}\\
\cos\Theta_{j2,j3}&=&\cos\Theta_{i,j2}\cos\Theta_{i,j3}
+\sin\Theta_{i,j2}\sin\Theta_{i,j3}\cos(\Phi_{i,j2}-\Phi_{i,j3}),
\la{eq325_3g}
\end{eqnarray} 
where one has to insert:
\beqn
\sin(\Phi_{i,j2}-\Phi_{i,i2})&=&\mbox{\hspace{-.3cm}}
\frac{\sin\Theta_{j2,i2}}{\sin\Theta_{i,i2}}\sin\Phi_{j2,i2},\la{eq325_3c}\\
\cos(\Phi_{i,j2}-\Phi_{i,i2})&=&\mbox{\hspace{-.3cm}}
\frac{\sin\Theta_{i,j2}\cos\Theta_{j2,i2}
-\cos\Theta_{i,j2}\sin\Theta_{j2,i2}\cos\Phi_{j2,i2}}{\sin\Theta_{i,i2}}.
\la{eq325_3d}
\eeqn
\end{appendix}

\vspace{0.5cm}

\begin{table}[h]\hspace{-.5cm}
\begin{tabular}{|lc|lc|}
\hline
\multicolumn{4}{|c|}{{CP, $\Gamma_{\alpha}=0$}}\\ \hline
\multicolumn{4}{|c|}{s:  $P$, $\Sigma^1_P$, $\Sigma^{11}_P$, $\Sigma^{22}_P$, 
$\Sigma^{33}_P$} \\
\multicolumn{4}{|c|}{a:  $\Sigma^3_P$, $\Sigma^{13}_P$, 
$\Sigma^{31}_P$} \\ 
\multicolumn{4}{|c|}{${\Sigma^2_P}=0$ , 
${ \Sigma^{12}_P}=0={\Sigma^{21}_P}$, 
${\Sigma^{23}_P}=0={\Sigma^{32}_P}$}\\
\hline\hline
 &\mbox{$\not\!\!\!\!{\mbox{C}\mbox{P}}$}, $\Gamma_{\alpha}=0$& 
&${\rm CP}, \Gamma_{\alpha}\neq 0$ \\ \hline
s: & $P$, $\Sigma^1_P$, ${ \Sigma^2_P}$, $\Sigma^{11}_P$, 
$\Sigma^{22}_P$, $\Sigma^{33}_P$, ${ \Sigma^{12}_P}$, 
${ \Sigma^{21}_P}$ & 
s: &
$P$, $\Sigma^1_P$, $\Sigma^{11}_P$, $\Sigma^{22}_P$,
$\Sigma^{33}_P$, ${ \Sigma^{23}_P}$, ${ \Sigma^{32}_P}$
\\
a: & $\Sigma^3_P$, $\Sigma^{13}_P$, $\Sigma^{31}_P$, 
${ \Sigma^{23}_P}$, ${ \Sigma^{32}_P}$ &
a: & ${ \Sigma^2_P}$, $\Sigma^3_P$, 
${ \Sigma^{12}_P}$, ${ \Sigma^{21}_P}$, 
$\Sigma^{13}_P$, $\Sigma^{31}_P$\\
\hline
\end{tabular}
\caption{ 
The forward--backward symmetry (s) and antisymmetry (a)
of all terms of the production spin--density 
matrix (\re{eq4_4h}) of Majorana fermions for unpolarized or longitudinally
polarized $e^+ e^-$ beams
with ($\Gamma_{\alpha}\neq 0$)
and without ($\Gamma_{\alpha}= 0$) consideration of the width of the 
exchanged particles $\alpha$ for CP conservation (CP) or CP violation 
( \mbox{$\not\!\!\!\!{\mbox{C}\mbox{P}}$}).  
\label{tab_-1}}
\end{table}
\vspace{3cm}
\begin{table}[h]\vspace{-3cm}
\hspace*{1.5cm}
\begin{tabular}{|l|c|c||c|c|}\hline
 & \multicolumn{2}{c||}{{Dirac fermions}} & 
\multicolumn{2}{c|}{{Majorana fermions}}\\
Decay distrib.
& \multicolumn{1}{|c|}{CP}  & \mbox{$\not\!\!\!\!{\mbox{C}\mbox{P}}$} 
& \multicolumn{1}{|c|}{CP} & \mbox{$\not\!\!\!\!{\mbox{C}\mbox{P}}$} \\ \hline
{energy } & $\Sigma^{3}_{P}$ &
$\Sigma^{3}_{P}$& -- &
$\Sigma^{3}_{P}$\\
{opening angle} & $\Sigma^{3}_{P}$ &
$\Sigma^{3}_{P}$& -- & $\Sigma^{3}_{P}$\\
{angular} & $\Sigma^{1,2,3}_{P}$ & $\Sigma^{1,2,3}_{P}$ & 
$\Sigma^{1,2,3}_{P}$ & $\Sigma^{1,2,3}_{P}$ \\
{siamese angle} & $\Sigma^{11,22,33}_{P}$ & $\Sigma^{ab}_{P}$ & 
$\Sigma^{11,22,33}_{P}$ & 
$\Sigma^{ab}_{P}$  \\ \hline
\end{tabular}\par\hspace*{1.7cm}
\parbox{9cm}{
\caption{For the energy and the different angular distributions
of the decay fermions in the lab frame  
the contributing polarizations and spin--spin correlations are specified. 
For CP conservation (CP) the polarization dependence is different 
for Majorana and Dirac fermions. If CP is violated 
( \mbox{$\not\!\!\!\!{\mbox{C}\mbox{P}}$})
all distributions
depend on spin correlations. \label{tab_1}    }
}      
\end{table}

\begin{figure}
\begin{center}
{\setlength{\unitlength}{1cm}
\begin{picture}(12,5)
\put(-.3,-1.2){\includegraphics{fig.1a}}
\put(-.5,2.5){\small$f(\vec{p}_1,\lambda_1)$}
\put(5.6,2.7){CPT}
\put(3.4,1.9){\small$\bar{f}(\vec{p}_2,\lambda_2)$}
\put(2.8,2.5){\small$\Theta$}
\put(1.8,3.7){\small $f_{i}(\vec{p}_i,\lambda_i)$}
\put(.9,1.1){\small $\bar{f}_{j}(\vec{p}_j,\lambda_j)$}
\put(11,2.5){\small$f(-\vec{p}_1,-\lambda_2)$}
\put(7.1,1.9){\small$\bar{f}(-\vec{p}_2,-\lambda_1)$}
\put(9.4,2.7){\small$\pi-\Theta$}
\put(9.8,3.7){\small $\bar{f}_{i}(\vec{p}_i,-\lambda_i)$}
\put(8.9,1.1){\small $f_{j}(\vec{p}_j,-\lambda_j)$}
\put(8.9,-.2){\small ${\cal R}_2(\pi)$}
\put(-.5,4){\small a)}
\end{picture}}\par\vspace{-1cm}
{\setlength{\unitlength}{1cm}
\begin{picture}(12,5)
\put(-.3,1){\includegraphics{fig.1b}}
\put(-.5,2.5){\small$f(\vec{p}_1,\lambda_1)$}
\put(5.8,2.7){CP}
\put(3.4,1.9){\small$\bar{f}(\vec{p}_2,\lambda_2)$}
\put(2.8,2.5){\small$\Theta$}
\put(1.9,3.6){\small $f_{i}(\vec{p}_i,\lambda_i)$}
\put(.6,.9){\small $\bar{f}_{j}(\vec{p}_j,\lambda_j)$}
\put(11.2,2.5){\small$\bar{f}(\vec{p}_2,-\lambda_1)$}
\put(7.2,1.9){\small$f(\vec{p}_1,-\lambda_2)$}
\put(9.6,1.8){\small$\pi-\Theta$}
\put(10.6,3.8){\small $f_{j}(-\vec{p}_j,-\lambda_j)$}
\put(8.6,.9){\small $\bar{f}_{i}(-\vec{p}_i,-\lambda_i)$}
\put(-.5,4){\small b)}
\end{picture}}\par\vspace{-1.2cm}
\end{center}
\caption{Production processes in the CMS under a) CPT
followed by a rotation ${\cal R}_2(\pi)$
and b) CP transformation. 
} \la{fig_1}
\vspace{1.5cm}
\begin{picture}(10,5)
\put(-.9,-.5){\includegraphics{fig.2}}
\put(2.6,3.4){ $\vec{p}_1$}
\put(8,.7){\small $\vec{p}_2$}
\put(7.7,2.4){\small $\vec{p}_i$}
\put(1.8,2.4){\small $\vec{p}_j$}
\put(7.1,1.5){\small $\Theta$}
\put(11.7,1.3){\small $\Theta$}
\put(11.3,.8){\small $\vec{p}_1$}
\put(7.9,3.5){\small $\vec{p}_{i3}$}
\put(10.4,5.5){\small $\vec{p}_{i2}$}
\put(10.1,-.3){\small $\vec{p}_{i1}$}
\put(8.5,5){\small $\Theta_{i2,i3}$}
\put(12,3.4){\small $\Theta_{i,i2}$}
\put(10.1,3){\small $\Theta_{1,i2}$}
\end{picture}\par\vspace{.2cm}
\caption{Definition of momenta and polar angles in the lab system.
The indices of the angles denote the plane covered by the corresponding 
momenta, the first index denotes the corresponding polar axis. 
The momenta and angles in the decay of $f_j$ are denoted
analogously.}
\label{fig_2}       
\end{figure}

\end{document}